\documentclass[11pt, oneside]{article}   	% use "amsart" instead of "article" for AMSLaTeX format
\usepackage{geometry}                		% See geometry.pdf to learn the layout options. There are lots.
\usepackage{graphicx}				% Use pdf, png, jpg, or eps§ with pdflatex; use eps in DVI mode
\usepackage{amssymb}
\usepackage{amsmath}
\usepackage{cite}
\usepackage{xcolor}

\usepackage{booktabs}
\usepackage{siunitx}
\usepackage{float}
\usepackage{authblk}
\usepackage{multicol}

\usepackage{babel}

\usepackage{lipsum}

\usepackage{nameref}
\usepackage{varioref}
\usepackage{hyperref}
\usepackage{cleveref}

\newcommand{\bSigma}{\mathbf{\Sigma}}
\newcommand{\bPsi}{\mathbf{\Psi}}
\newcommand{\bS}{\mathbf{S}}
\newcommand{\bone}{\mathbf{1}}
\newcommand{\bU}{\mathbf{U}}
\newcommand{\bu}{\mathbf{u}}
\newcommand{\by}{\mathbf{y}}

\newcommand{\bV}{\mathbf{V}}
\newcommand{\bI}{\mathbf{I}}
\newcommand{\bX}{\mathbf{X}}
\newcommand{\bx}{\mathbf{x}}

\newcommand{\indep}{\rotatebox[origin=c]{90}{$\models$}}

\newtheorem{theorem}{Theorem}[section]

\newtheorem{proposition}[theorem]{Proposition}

\def\keywords{\vspace{.5em}
{\textit{Keywords}:\,\relax%
}}

\begin{document}

\sloppy

\title{Bayesian Sensitivity Analysis for Missing Data Using the E-value}

\author{Wu Xue \\ Department of Statistics \\ The George Washington University \\ \href{marcoxue516@gwu.edu}{marcoxue516@gwu.edu} 
   \and Abbas Zaidi  \\ Statistics \& Privacy  \\Facebook \\ \href{abbaszaidi@fb.com}{abbaszaidi@fb.com} }

\date{}
							% Activate to display a given date or no date
\maketitle

\begin{abstract}

\noindent \textit{Sensitivity Analysis} is a framework to assess how conclusions drawn from missing outcome data may be vulnerable to departures from untestable underlying assumptions. We extend the E-value, a popular metric for quantifying robustness of causal conclusions,  to the setting of missing outcomes. With motivating examples from partially-observed \textit{Facebook} conversion events, we present methodology for conducting Sensitivity Analysis at scale with three contributions. First, we develop a method for the Bayesian estimation of sensitivity parameters leveraging noisy benchmarks (e.g., aggregated reports for protecting unit-level privacy); both empirically derived subjective and objective priors are explored. Second, utilizing the Bayesian estimation of the sensitivity parameters we propose a mechanism for posterior inference of the E-value via simulation. Finally, closed form distributions of the E-value are constructed to make direct inference possible when posterior simulation is infeasible due to computational constraints. We demonstrate gains in performance over asymptotic inference of the E-value using data-based simulations, supplemented by a case-study of Facebook conversion events. 

\end{abstract}


\section{Introduction}\label{sec: int}

%\blue{
%Outline of introduction:
%}
%\begin{itemize}
%
%\color{blue}
%\item State the problem that we are facing and propose solutions, bayesian sensitivity analysis. Including:
%\begin{itemize}
%\item Accommodate variance differences when comparing debiaser and AEM
%
%\item Give uncertainty estimate of E-values, using empirical Bayesian method
%
%\end{itemize}
%
%\item Review sensitivity analysis in causal inference and missing data problem, explain how E-values could be used in missing data problems as well.
%
%\item Review pattern mixture model and Bayesian pattern mixture model. Make clear the relationship between sensitivity parameter and E-value.
%
%\item Describe briefly proposed solutions in this work. Make clear selection of prior could make material difference.
%
%\item Outline structure of the paper.
%
%\end{itemize}

An increasing number of statistical methods have been developed to garner meaningful inference from missing data, an inevitability in a multitude of applications.
%Facebook is currently tackling 
%\textit{Signal Loss} which describes an array of policies under which event level conversion data (e.g. purchases, add-to-cart) may be lost: a typical missing data problem. % Wu, this is how I would re-write it %.
%This data loss materially impacts Facebook's Ads business %Wu, this goes beyond Delivery of Ads or their Measurement, so you want to make a more general statement %,
%along several axes, including how ads are fairly auctioned, delivered and eventually measured for effectiveness. %Wu, you generally want to avoid using internal designations like 'Debiaser' -- best to abstract away to scientific naming conventions.%
These include ad-hoc solutions relying on strong assumptions about the missingness mechanism that are implausible in practice, such as the analysis of complete cases. Model-based approaches such as multiple imputation (MI) and inverse probability weighting (IPW) are proposed to deal with missing data under less restrictive assumptions. Both techniques 
%
%To mitigate these impacts, Facebook relies on \textit{Inverse Probability Weighting} (IPW) of observable data which
leverage the \textit{Missing at Random} (MAR) assumption on the missingness mechanism which states that missingness is \textit{random} conditional on the observed features. 

While it is reasonable to assume MAR, the possibility of data being \textit{Missing Not at Random} (MNAR) can never be fully excluded; When data is potentially MNAR, methods such as IPW would yield inconsistent (and non-identified) estimates. However, since the missingness mechanism is inherently a statement about \textit{unobserved} data, direct validation is not possible; this necessitates indirect evaluation to understand the impacts of MNAR on conclusions. One framework for doing so is  \textit{Sensitivity Analysis} \cite{RoseR83}.%WU: ADD A REFERENCE TO THE ROSENBAUM PAPER%. The objective of this umbrella of methods is to substantively induce MNAR and examine the extent to which conclusions are expected to change as a consequence.

%Because of the material effect of Debiaser on Ads Delivery system, to develop solid Debiaser validation framework is of vital importance, and one of the fundamental issue in Debiaser validation is to perform sensitivity analysis to check the extent to which Debiaser is away from the approximate ground truth.% -- You don't need this here at the moment since its detail more about the Debiaser so we can condense it down.

%For example, if values are missing completely at random (MCAR), we may ignore observations with missing data since the cause of the missing data are unrelated with the data. If values are MAR, which means the probability of being missing is not related to the missing data, but can be accounted for by observed variables, we may apply methods such as IPW to reduce bias when only complete cases are used.  When neither MCAR nor MAR holds, it would be called missing not at random (MNAR), which requires much more complicated methods to deal with and sensitivity analysis needs to be performed to see how sensitive the results are under different mechanisms. 

A myriad of research \cite{KaciR14, McCaGL07} has been undertaken on Sensitivity Analysis for missing data problems. The representative monograph \cite{LittDC12} introduces several methods including Pattern Mixture Model and Selection Model approaches as the dominant families. Related techniques have also been used in Causal Inference \cite{CarnHH16, VandD17, FranDF19} with some applications that may be extended to missing data problems. For example, \cite{VandD17} proposed the E-value to show how robust causal effect estimates are against unmeasured confounding. Treating the MAR assumption as a form of ignorability, we will discuss how the E-value is applicable to missing data problems \cite{MoleTJ04}.%ADD A REFERENCE HERE%.

In order to effectively utilize the E-value to understand the implications of MAR violations, estimating its uncertainty is crucial; an area that is both underdeveloped and necessary. 
%In the mean time, selecting ranges of sensitivity parameters in Sensitivity Analysis is infeasible at scale. Therefore, we aim to estimate uncertainty of E-value under the Bayesian paradigm.  
%IN THIS PAPER WE PROPOSE ETC ETC %Wu, you can add this in and refine using what I have written out so far%
In this paper we propose to cast the estimation and uncertainty quantification of the E-value as a \textit{Bayesian Inference} problem. Conceptually, these results are rooted in the Bayesian estimation of the \textit{sensitivity parameter} using a combination of \textit{noisy benchmarks} and \textit{prior information}. These benchmarks  are becoming increasingly common in applications where aggregated information is used as a means of protecting privacy at the unit-of-observation level (the objective of inference). Examples of these include but are not limited to Google's \textit{Privacy Sandbox} \cite{GeraKK21},  %\text{Facebook's Aggregated Event Measurement (AEM)}
\textit{differentially private aggregates} \cite{LiuKKO21} or Facebook's \textit{Aggregated Event Measurement} (AEM) system. Under this overarching theme, the Bayesian estimation of the sensitivity parameter induces a posterior distribution for the E-value that can either be approximated using simulation \cite{GelfS90} or be analytically determined under a series of testable assumptions where simulation is infeasible. % WU ADD A REFERENCE TO MCMC HERE%. Analytic approximations for the posterior distribution of the E-value under mild assumptions are also provided.

The remainder of the paper proceeds as follows: Section \ref{sec: back} provides a brief overview of Sensitivity Analysis for inference from missing data along with a summarization of the contributions of this paper. In Section \ref{sec: method}, a Bayesian approach using either an objective or an empirical subjective prior for the sensitivity parameter is proposed with an inference scheme for the E-value. Section \ref{sec: simu} presents a simulation study to evaluate the performance of the proposed method. Finally, this technique is applied to the validation of IPW estimates from \textit{Facebook} data in Section \ref{sec: data}; this exercise is motivated by methodology used on the platform for estimating aggregates from partially missing outcomes. We conclude the paper in section \ref{sec: disc} with a discussion of our findings as motivation for future work.

\section{Background} \label{sec: back}

%Sensitivity Analysis is the study of how statistical findings or conclusions are affected by changing key assumptions in empirical data analysis. In the field of missing data Sensitivity Analysis typically refers to approaches that assess the sensitivity of estimates under the MNAR assumption than the MAR assumption, a common assumption used in missing data problems.  NOTE: Wu, you don't need this in the background%

Sensitivity Analysis techniques for the assessment of MAR violations (i.e., the missingness mechanism is really MNAR) fall into the two dominant categories: \textit{Pattern Mixture Models} and \textit{Selection Models}. 
% NOTE: Avoid using notation here; you want to write this in as close to 'natural language statements' as possible. Basically use words rather than symbols.
Both approaches factorize the joint distributions of the measurement (i.e., the outcomes) and the missingness, albeit a bit differently. Pattern Mixture Models factorize the joint distribution into the conditional distribution of the measurement given missingness and the marginal distribution of the latter \cite{Litt93, Litt94, LittW96} while Selection Models reverse this decomposition \cite{DiggK94, Heck76, HeytTE92}. Both families of techniques present benefits and challenges which signal suitability for our motivating application.

Selection Models are appealing since their focus is on the estimation of the conditional distribution of the missingness. This conditional representation can accommodate auxiliary information easily. However, model checking for this setting is an underdeveloped area with best practice suggesting flexible approaches. The robustness afforded by the added flexibility induces higher variance (and therefore inefficiency) in resulting estimates. Another non-trivial constraint is setting sensitivity parameters, particularly with continuous measurements since their interpretation on various scales may not be transparent within the aforementioned conditional distribution.
%a parametric model for $R$ given $Y$ is required; 
By comparison, Pattern Mixture Models incorporate assumptions  
about the missingness mechanism via the sensitivity parameter. These are directly interpretable as the differences in the conditional expectations of the measurement by missingness status\cite{Litt08}. This lends itself to ease of interpretation and therefore ease in determining plausible values for said parameters. The simplification here comes at the expense of challenges in incorporating auxiliary information. Furthermore, in certain settings, this setup may induce more complexity in derivation of estimators (necessitating additional simplifying assumptions). %NOTE: Wu, this is not strictly true, i.e. you can potentially incorporate a model for R|Y into the pattern mixture model as well. You should flesh out their differences a bit more thoroughly. Take a look at the book chapter I sent you before your internship. That might help lay this out a bit more.
It warrants mention that any assumptions about the missingness mechanism for either approach are not directly verifiable from the data. For the application of interest, Pattern Mixture Models are uniquely suited \cite{RoseR83}; employing privacy motivated aggregates of the measurement conditional on the missingness, enable substantive inference on the sensitivity parameters.

Earlier work on Sensitivity Analysis using Pattern Mixture Models relied on expert knowledge to select sensitivity parameter values, i.e., on average, the extent to which we expect the identifying MAR assumption to be violated. To circumvent an inappropriate selection, Bayesian methods have been applied in Sensitivity Analysis to weaken the reliance on untestable assumptions. \cite{GaskDM16} proposed using Bayesian shrinkage on the mean and dependence parameter to share information across different missingness patterns. \cite{KaciR14} introduced a Bayesian approach to analyze outcomes from the exponential distribution family with missing values that are MNAR.  \cite{MasoGG17} proposed a Bayesian approach to deal with missing data when estimating causal effects in randomized clinical trials. Although there is a rich literature on using Bayesian approaches to assess missing data, inference  on the sensitivity parameter is largely limited to subject matter expertise driven priors. Given the scale of our applications for Facebook data, relying on expert information to elicit priors or choosing prior hyper-parameters manually is infeasible.

In this work, we propose an approach to Sensitivity Analysis using Bayesian estimation of the sensitivity parameters from noisy, aggregated data. We apply either empirically derived subjective priors (when noisy but collectively useful data is available) or objective priors (when high quality data with strong unit information is available); both techniques allow automation and are suited for scalability as a result. These Bayesian estimates can be used to infer the E-value of \cite{VandD17} to summarize sensitivity to MAR violations with uncertainty induced via the sensitivity parameter. Furthermore, we derive analytical forms of the distribution function of the E-value based on the posterior distribution of the sensitivity parameter. These provide further possibilities for scalable Sensitivity Analysis where simulation based posterior inference may not be viable.

%We also incorporate the conception of E-value in \cite{VandD17} to give more interpretable results. % Wu, two comments here. First, we are not just looking at objective priors (since we are also looking at empirical subjective priors so you need to talk about that). Second you need to justify, again, not mathematically, but in natural language why we can use objective priors here.

% Wu, this should be in the intro.
%The rest of the paper proceeds as follows. In METHODOLOGY, a Bayesian approach with objective priors for sensitivity parameter is proposed to conduct sensitivity analysis. SIMULATION section presents a simulation study to evaluate the performance of the proposed method. In DATA APPLICATION, the proposed method is applied to the validation process of IPW estimates; a part of the Facebook project. DISCUSSION concludes the paper. 

\section{Methodology} \label{sec: method}

\subsection{Notation and Assumptions}
For units of observation $i = 1, \ldots, n$, let $Y_{i}$ and $\bX_{i}$ denote the continuous outcome and covariates respectively. Furthermore, let $R_{i}$ denote the missingness such that $R_i = 1$ if $Y_i$ are observed and $R_i = 0$ otherwise. We will focus on the estimation of the population mean, $\mathbb{E}[Y] = \mu$.

If the missingness mechanism is \textit{Missing Completely at Random}, i.e., MCAR ($Y_i ~ \indep  ~ R_i$), one can estimate $\mu$ by $\hat{\mu} = \left( \sum_{i = 1}^n R_i \right)^{-1} \sum_{i=1}^n R_i Y_i$ consistently; in a myriad of applied settings this assumption is implausible. A more likely scenario assumes that the mechanism is \textit{Missing at Random}, i.e., MAR ($Y_i ~ \indep ~ R_i | \bX_i$); 
%$\mu$; 
Under this variation on the missigness mechanism, the population mean can be estimated by techniques including IPW or MI. We will focus on the IPW estimator, i.e., $\hat{\mu} = n^{-1} \sum_{i=1}^n R_iY_i / \pi(\bX_i) $ where $\pi(\bx) = \text{pr} (R = 1 | \bX = \bx) $ is the true propensity score. In practice, true propensity scores are unknown but estimable; with a consistent estimate of the propensity score, $\mu$ can be consistently estimated. 

Unfortunately, any statement about the missingness mechanism is a statement about \textit{unknown unknowns} and so MNAR can never be fully excluded from possibility. Let $\delta$ denote the sensitivity parameter in the underlying Pattern Mixture Model which represents the degree to which MNAR is induced. This parameter is usually selected based on substantive assumptions. 

An alternative is to estimate it as $\hat{\delta} = \hat{\mu} - \mu$; In our motivating applications at Facebook (we briefly touch on others in section \ref{sec: int}), $\mu$ may be observed but is contaminated by noise (e.g., for the purposes of protecting privacy). We focus on the scenario where there are \textbf{two} distinct sources for $\mu$ and therefore, two estimates of the sensitivity parameter.

%$H$ denote the hold-out status with $1$ for being in the hold-out group and $0$ for otherwise, where being hold-out means the subject is fed with random advertisements. For each subject in the hold-out group, an estimated propensity score for being opt-in user, where being opt-in means that the subject consents to share information with Facebook, i.e. observable by Facebook. Let $\hat{\mathbb{E}(Y)}$ denote the IPW estimator for $\mathbb{E}(Y)$, and based on pattern mixture model, we define sensitivity parameter to be $\hat{\delta} = \hat{\mathbb{E}(Y)} - \mathbb{E}(Y)$. There should exist more than one source of obtaining ground truth $\mathbb{E}(Y)$ and in our case we consider two sources of obtaining ground truth for $\mathbb{E}(Y)$. 

Let $\boldsymbol\delta_j = (\delta_{j1}, \delta_{j2})$ denote the sensitivity parameter estimates for groups $j = 1, \ldots, m$ in the population that are leveraged jointly to learn $\delta$. Let the \textit{likelihood} function for $m$ groups, $f(\boldsymbol\delta_j|\delta\mathbf{1}, \bSigma)$ be a bi-variate normal distribution with mean vector $\delta\mathbf{1}$ and covariance matrix $\bSigma$. Under the Bayesian paradigm, we specify both empirical subjective and objective \textit{priors} over $(\delta, \bSigma)$ to conduct inference.

\subsection{Empirical Subjective Prior}
We choose the Normal-Inverse-Wishart distribution as the form of the subjective prior over $(\delta, \bSigma)$,
\begin{align*}
\pi(\delta | \bSigma) &\sim \mathrm{N}(\delta_0, \phi_0 ), ~ \text{where} ~ \phi_0 = \left( \bone ^{\prime}\bSigma^{-1} \bone \right)^{-1},\\
\pi(\bSigma) &\sim \mathrm{IW}(\bPsi, \nu).
\end{align*}
Then the joint density $f(\boldsymbol\delta_1, \ldots, \boldsymbol\delta_m, \delta, \bSigma)$ is given by
\begin{align}
f(\boldsymbol\delta_1, \ldots, \boldsymbol\delta_m, \delta, \bSigma) = \frac{1}{(2\pi)^m |\bSigma |^{m/2}}e^{-\frac{1}{2} \text{tr}(\bS_0\bSigma^{-1})}\frac{1}{\sqrt{2\pi \phi_0}}e^{-\frac{(\delta - \delta_0)^2}{2\phi_0}}\cdot
\end{align}
$$
\frac{|\bPsi|^{\nu/2}}{2^{\nu}\Gamma_2(\nu/2)}|\bSigma|^{-\frac{\nu+3}{2}}e^{-\frac{1}{2}\text{tr}(\bPsi\bSigma^{-1})},
$$

where $\bS_0 = \sum_{j=1}^m (\boldsymbol\delta_j -  \delta \bone) (\boldsymbol\delta_j -  \delta \bone)^{\prime}$. 
\newline

In order to learn the \textit{optimal} settings for the hyper-parameters $\bPsi$, $\delta_{0}$ and $\nu$, we derive the marginal likelihood over $\boldsymbol\delta_1, \ldots, \boldsymbol\delta_m$ which is given by integrating over $\delta$ and $\bSigma$,

\begin{align*}
m(\boldsymbol\delta_1, \ldots, \boldsymbol\delta_m) &= \iint f(\hat{\delta}_1, \hat{\delta}_2, \delta, \bSigma) \text{d} \delta \text{d}\bSigma \\
&=
\iint \frac{1}{(2\pi)^m |\bSigma |^{m/2}}e^{-\frac{1}{2} \text{tr} (\bS_0\bSigma^{-1})}\frac{1}{\sqrt{2\pi \phi_0}}e^{-\frac{(\delta - \delta_0)^2}{2\phi_0}}\cdot \\ &\frac{|\bPsi|^{\nu/2}}{2^{\nu}\Gamma_2(\nu/2)}|\bSigma|^{-\frac{\nu+3}{2}}e^{-\frac{1}{2} \text{tr}(\bPsi\bSigma^{-1})} \text{d} \delta \text{d}\bSigma.
%&= 
%h(\delta_0, \lambda, \bPsi, \nu)\iint |\bSigma|^{-\frac{n + \nu+3}{2}} \exp\left[ -\frac{1}{2} \text{tr} \bSigma^{-1}\left\{ \bPsi + \sum_{i=1}^n (\boldsymbol\delta_i -  \delta \bone) (\boldsymbol\delta_i -  \delta \bone)^{\prime} + \lambda(\delta\bone - \delta_0\bone)(\delta\bone - \delta_0\bone)^{\prime} \right\} \right] \\
%&=
%h(\delta_0, \lambda, \bPsi, \nu)\iint |\bSigma|^{-\frac{n + \nu+3}{2}} \exp\left[ -\frac{1}{2} \text{tr} \bSigma^{-1}\left\{ \bPsi + \bS + n(\bar{\boldsymbol\delta} - \delta \bone)(\bar{\boldsymbol\delta} - \delta \bone)^{\prime} + \lambda(\delta\bone - \delta_0\bone)(\delta\bone - \delta_0\bone)^{\prime} \right\} \right] \\
\end{align*}

Leveraging conjugacy, the marginal likelihoood can be derived using the \textit{posterior} distribution over  $\delta$ and $\bSigma$,
\begin{align*}
\pi(\delta, \bSigma | \boldsymbol\delta_1, \ldots, \boldsymbol\delta_m) 
&\propto
\exp\left[ -\frac{1}{2} \text{tr} \bSigma^{-1}\left\{ \bPsi + \sum_{j=1}^m (\boldsymbol\delta_j -  \delta \bone) (\boldsymbol\delta_j -  \delta \bone)^{\prime} + (\delta\bone - \delta_0\bone)(\delta\bone - \delta_0\bone)^{\prime} \right\} \right] \\
&\propto
\exp\left[ -\frac{1}{2} \text{tr} \bSigma^{-1}\left\{ \bPsi + \bS + m(\bar{\boldsymbol\delta} - \delta \bone)(\bar{\boldsymbol\delta} - \delta \bone)^{\prime} + (\delta\bone - \delta_0\bone)(\delta\bone - \delta_0\bone)^{\prime} \right\} \right].
\end{align*}

We have,
\begin{align*}
\pi(\delta, \bSigma | \boldsymbol\delta_1, \ldots, \boldsymbol\delta_m)  \sim \mathrm{NIW}\left(\tilde{\delta}, \tilde{\Psi}, \tilde{\nu}\right),
\end{align*}
where 
\begin{align*}
\tilde{\delta} &= \frac{  \delta_0  + (m/2) \bar{\boldsymbol\delta}^{\prime}\bone}{ m + 1}, \\
%\tilde{\lambda} &= \lambda + n \\
\tilde{\bPsi} &= \bPsi + \bS + \frac{ m}{m+1}(\bar{\boldsymbol\delta} - \delta_0 \bone)(\bar{\boldsymbol\delta} - \delta_0 \bone)^{\prime}, \\
\tilde{\nu} &= \nu + m,  \\
\bS &= \sum_{j=1}^m (\boldsymbol\delta_j-  \bar{\boldsymbol\delta}) (\boldsymbol\delta_j -  \bar{\boldsymbol\delta}) ^{\prime}.
\end{align*}
%where $h(\delta_0, \lambda, \bPsi, \nu) = \frac{|\bPsi|^{\nu/2}}{2^{\nu + n + 1/2}\Gamma_2(\nu/2) \pi ^{n +1/2}} $, $\bS = \sum_{i=1}^n (\boldsymbol\delta_i -  \bar{\boldsymbol\delta}) (\boldsymbol\delta_i -  \bar{\boldsymbol\delta})^{\prime}$.
Then the marginal likelihood is the ratio of the joint density $f(\boldsymbol\delta_1, \ldots, \boldsymbol\delta_m, \delta, \bSigma)$ to the posterior distribution $\pi(\delta, \bSigma | \boldsymbol\delta_1, \ldots, \boldsymbol\delta_m)$, 
\begin{align} \label{equation: likelihood}
m(\boldsymbol\delta_1, \ldots, \boldsymbol\delta_m) 
&=
    (2\pi)^{-m} \frac{|\bPsi|^{\nu/2} }{2^{\nu} \Gamma_2(\nu/2)}\frac{2^{\tilde{\nu}} \Gamma_2(\tilde{\nu}/2)}{|\tilde{\bPsi}|^{\tilde{\nu}/2} } \nonumber \\
    &=
    \frac{1}{\pi^m} \frac{|\bPsi|^{\nu/2}}{|\tilde{\bPsi}|^{\tilde{\nu}/2}}\frac{ \Gamma_2(\tilde{\nu}/2)}{\Gamma_2(\nu/2)}.
\end{align}

Taking the negative logarithm of the marginal likelihood function yields the objective function with respect to $\delta_0,  \bPsi$ and $\nu$ that can be minimized to learn the optimal parameter settings for the subjective prior,

\begin{equation} \label{equation: optim}
\mathcal{L} (\delta_0, \bPsi, \nu) 
=
m \log(\pi)  - \frac{\nu}{2} \log |\bPsi| + \frac{\tilde{\nu}}{2} \log | \tilde{\bPsi} | + \log \frac{ \Gamma_2(\nu/2)}{\Gamma_2(\tilde{\nu}/2)}.
\end{equation}

\begin{proposition}\label{proposition: convex}
Let $\Psi \in \mathbb{R}^{d\times d}, d \in \mathbb{N}^{+}$ be a symmetric matrix. 
The objective function $\mathcal{L} (\delta_0, \bPsi, \nu)$  is convex with respect to $\Psi$ when $  \nu > C m $ where $C$ is some constant. Furthermore, when $\bPsi = \bPsi^{*}$, its optimal value, the objective function is convex in $\delta_0$ when the squared \textit{Mahalanobis} distance under \textbf{S} between $\bar{\boldsymbol\delta}$ and $\delta_0 \bone$ is bounded by some constant $K$ depending on $m$. For proof see Appendix \ref{sec:app}.

\end{proposition}

Let $(\bPsi^*, \nu^*, \delta_0^*)$ be the global minimizer of equation (\ref{equation: optim}). Substituting $(\bPsi^*, \nu^*, \delta_0^*)$ into the posterior distribution of $(\delta, \bSigma)$ and integrating out $\bSigma$ gives the marginal posterior distribution of $\delta$ \cite{Geis65},

\begin{align}
P(\delta) &\propto \bigg | (\delta - \delta_0^{*})^2 \bone \bone^{\prime} + \bPsi^* + \sum_{j = 1}^m (\boldsymbol\delta_j - \delta \bone)(\boldsymbol\delta_j - \delta \bone)^{\prime} \bigg | ^{-(m+ \nu^*)/2} \nonumber \\
&\propto
\left[ 1 + (m + 1)(\bar{\by} - \delta \bone)^{\prime} \bU^{-1} (\bar{\by} - \delta \bone) \right]^{-(m+ \nu^*)/2},
\end{align}
where 
\begin{align*}
\bU &= \bPsi^* + \bS + \frac{m}{m+1} (\bar{\boldsymbol\delta} - \delta_0^* \bone) (\bar{\boldsymbol\delta} - \delta_0^* \bone) ^{\prime}, \\
\bar{\by} &= \frac{m\bar{\boldsymbol\delta} + \delta_0^* \bone}{m +1}.
\end{align*}
Let $u = \bone^{\prime}\bU^{-1} \bar{\by}, z = \bone^{\prime}\bU^{-1} \bone$ and $w = \by^{\prime} \bU^{-1} \by$, the marginal posterior distribution of $\delta$ follows a \textit{generalized} Student's \textit{t}-distribution with $ m + \nu^* -1$ degrees of freedom,
\begin{align}
P(\delta) \propto \left[ 1 + \frac{ (m+1) z \left(\delta - \frac{u}{z} \right)^2}{1 + (m+1)w - (m+1)u^2z^{-1}} \right] ^{-(m+ \nu^*)/2}.
\end{align}

\subsection{Objective Prior}
Under the objective Bayesian umbrella, we choose the independent Jeffreys prior, $\pi_{IJ} = |\bSigma|^{-(p+1)/2}$. Since $p = 2$ in our motivating application, the independent Jeffreys prior has the form,
\begin{align}
\pi_{IJ}( \delta, \bSigma) = |\bSigma|^{-3/2}.
\end{align}

Then the joint density $f(\hat{\delta}_1, \hat{\delta}_2, \delta, \bSigma)$ is given by
\begin{align}
f(\hat{\delta}_1, \hat{\delta}_2, \delta, \bSigma) &= \frac{1}{(2\pi)^m |\bSigma |^{m/2}}e^{-\frac{1}{2} \text{tr}\{\sum_{j=1}^m (\boldsymbol\delta_j -  \delta \bone) (\boldsymbol\delta_j -  \delta \bone) ^{\prime}\bSigma^{-1} \} } |\bSigma|^{-3/2}  \nonumber \\ 
&=
\frac{1}{(2\pi)^m }e^{-\frac{1}{2} \text{tr} \{ \sum_{j=1}^m(\boldsymbol\delta_j -  \delta \bone) (\boldsymbol\delta_j -  \delta \bone) ^{\prime}\bSigma^{-1} \} } |\bSigma|^{-(m+3)/2},
\end{align}
%where $\bS = \sum_{i=1}^n(\hat{\delta}_{i1} - \delta, \hat{\delta}_{i2}-\delta)^{\prime}(\hat{\delta}_{i1} - \delta, \hat{\delta}_{i2}-\delta)$. a
and the marginal posterior distribution over $\delta$ is given by marginalizing over $\bSigma^{-1}$ \cite{Geis65},

\begin{align*}
P(\delta) &= \int f(\hat{\delta}_1, \hat{\delta}_2, \delta, \bSigma)  \text{d}\bSigma \\
&= 
\int \frac{1}{(2\pi)^m }e^{-\frac{1}{2} \text{tr} \{ \sum_{j=1}^m (\boldsymbol\delta_j -  \delta \bone) (\boldsymbol\delta_j -  \delta \bone) ^{\prime}\bSigma^{-1} \} } |\bSigma|^{-(m+3)/2} \text{d}\bSigma \\
&\propto
 \left| \sum_{j=1}^m (\boldsymbol\delta_j -  \delta \bone) (\boldsymbol\delta_j -  \delta \bone) ^{\prime} \right|^{-m/2}.
\end{align*}

Now let 
$$\bar{\boldsymbol\delta} = m^{-1}\sum_{j=1}^m\boldsymbol\delta_j \quad  \text{and} \quad \bS = \sum_{j=1}^m (\boldsymbol\delta_j -  \bar{\boldsymbol\delta}) (\boldsymbol\delta_j -  \bar{\boldsymbol\delta}) ^{\prime}, $$
and therefore,
\begin{align*}
\sum_{j=1}^m (\boldsymbol\delta_j -  \delta \bone) (\boldsymbol\delta_j -  \delta \bone) ^{\prime} = \bS + m(\bar{\boldsymbol\delta} - \delta \bone)(\bar{\boldsymbol\delta} - \delta \bone)^{\prime}.
\end{align*}
Recall that
\begin{align*}
\left| I + m(\bar{\boldsymbol\delta} - \delta \bone)(\bar{\boldsymbol\delta} - \delta \bone)^{\prime}  \bS^{-1} \right|  = 1 + m(\bar{\boldsymbol\delta} - \delta \bone)^{\prime} \bS^{-1}(\bar{\boldsymbol\delta} - \delta \bone).
\end{align*}
Define $u = \bone^{\prime} \bS^{-1}\bar{\boldsymbol\delta}, z = \bone^{\prime} \bS^{-1} \bone$ and $w = \bar{\boldsymbol\delta}^{\prime}\bS^{-1}\bar{\boldsymbol\delta}$, we have
\begin{align*}
P(\delta) 
&\propto
 \left[ 1 + \frac{mz\left(\delta - \frac{u}{z}\right)^2}{1 + mw - mu^2 z^{-1}} \right]^{-m/2}.
\end{align*}
Therefore, the marginal posterior distribution of $\delta$ in this setting is also a \textit{generalized} Student's \textit{t}-distribution with $m-1$ degrees of freedom. 
\newline
%\textcolor{red}{The relationship between the likelihood, objective prior and the posterior distributions is given in Figure...}

\begin{figure}[ht]
\centering
\includegraphics[width=0.8\textwidth]{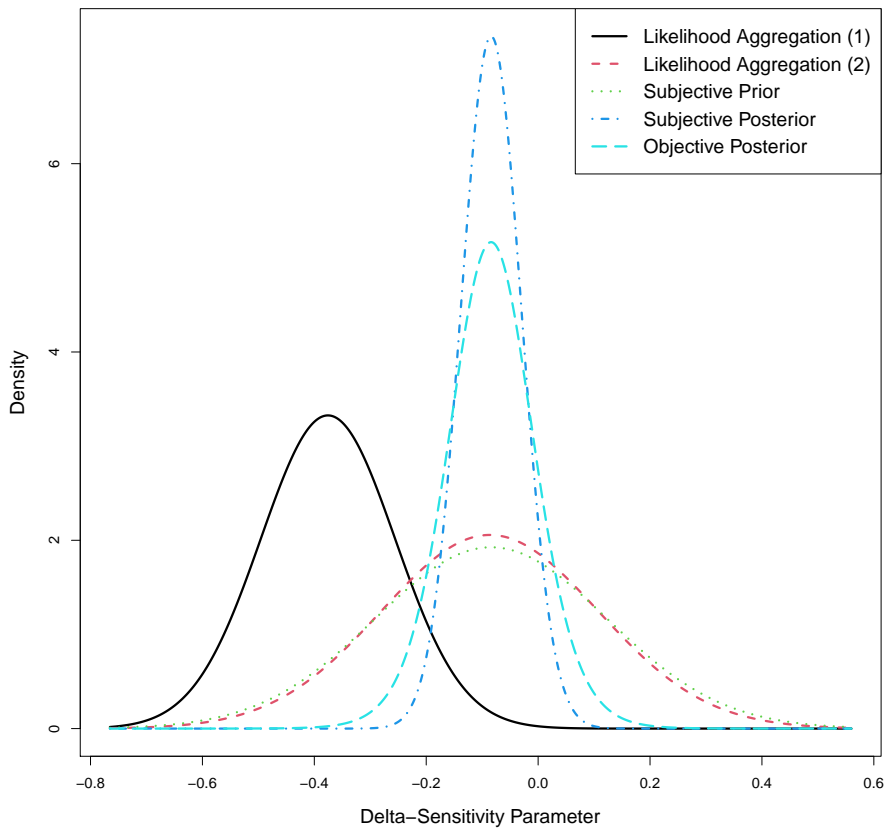}
\caption{The likelihood functions based on two noisy sources for $\mu$, the empirically motivated subjective and objective priors and the corresponding posterior distributions.}
\label{fig: density1}
\end{figure}

The relationship between the likelihood, priors and their corresponding posterior distributions is given in Figure \ref{fig: density1}.

\subsection{Bayesian Inference for the E-value}

The E-value was introduced in 
\cite{VandD17} to quantify the impacts of unmeasured confounding on the difference in continuous measurements. This technique rests on a \textit{standardized effect size}, i.e., a scaled difference between $\hat{\mu}$ and $\hat{\mu}_{\delta}$, where $\hat{\mu}_{\delta}$ denotes the estimate of the population mean incorporating the sensitivity parameter. This can be used to approximate the risk ratio which in turn yields the E-value. In this context, if there is no material impact from MAR violations, we expect the E-value to be \textit{statistically indistinguishable} from its reference value 1, i.e., there is no meaningful difference between $\mu$ and $\mu_{\delta}$.

For inference, the posterior distribution of the E-value can then be approximated by simulation  \cite{GelfHRS90, DiebR94} using the following formulation based on the standardized effect size.
%WU ADD A REFERENCE TO MCMC HERE%.
We decompose $\mu$ using the \textit{Law of Iterated Expectations},

\begin{align*}
\mu= \mathbb{P}(R = 1) \mathbb{E}(Y | R = 1) + \mathbb{P}(R = 0) \mathbb{E}(Y | R = 0).
\end{align*}

This is the foundation of the Pattern Mixture Model approach which decomposes the mean into the unobserved $\mathbb{E}(Y | R = 0)$ and observed $\mathbb{E}(Y | R = 1)$ components. For identification assume that 

\begin{align*}
\mathbb{E}(Y | R = 0)  =  \mathbb{E}(Y | R = 1) + \delta.
\end{align*}

Substituting this back into the decomposition of $\mu$ yields, 

\begin{equation}
\mu_{\delta} = \mathbb{P}(R = 1) \mathbb{E}(Y | R = 1) + \left\{1 - \mathbb{P}(R = 1)\right\}  \left\{ \delta + \mathbb{E}(Y | R = 1) \right\}.
\end{equation}

Let $\mu_{missing}$ denote the standardized effect size which can be calculated as,

\begin{align}
\mu_{missing} &= \frac{\mu_{\delta = \delta} - \mu_{\delta = 0} }{ \sqrt{ \text{Var}(Y) } }\nonumber \\
&=
 \frac{ \left\{1 - \mathbb{P}(R = 1)\right\}  \delta }{ \sqrt{ \text{Var}(Y) } },
 \end{align}
and the corresponding risk ratio (RR) can be approximated by 
\begin{align*}
RR \approx \exp( 0.91 \times \mu_{missing} ),
\end{align*}
and E-value can be obtained as,
\begin{align*}
\text{E-value} = RR + \sqrt{RR  (RR - 1)}.
\end{align*}

In addition to inference via posterior simulation, under certain assumptions, we can also approximate analytic distribution functions of the E-value under the framework in \cite{VandD17}. For brevity, let $V$ denote the E-value; Using the formulation presented earlier, we have the following theorems on the distribution of $V$, $f_{V}(v)$ (For proofs see Appendix \ref{sec:app}).

\begin{theorem} \label{theorem: evalue1}
Suppose that $\mathbb{P} ( R = 1) = p$ and $\text{Var}(Y) = \sigma_Y$ are known, and that $\delta$ follows a normal distribution $\mathrm{N}(\eta, \tau^2)$. Then the 
density function of $V$ is 
\begin{equation}
f_V(v) = 
\begin{cases}
\frac{1}{\sqrt{2 \pi} \sigma_{RR}} \exp \left\{ - \frac{ \left( \ln \frac{v^2}{2v-1} - \mu_{RR} \right)^2 } { 2 \sigma_{RR}^2} \right\} \left\{ \frac{1}{v} - \frac{1}{v} \frac{1}{2v-1} \right\} & \text{if $RR > 1$}, \\
\frac{1}{\sqrt{2 \pi} \sigma_{RR}} \exp \left\{ - \frac{ \left( \ln \frac{2v-1}{v^2} - \mu_{RR} \right)^2 } { 2 \sigma_{RR}^2} \right\} \left\{ \frac{1}{v} - \frac{1}{v} \frac{1}{2v-1} \right\} & \text{if $0 < RR < 1$} ,
\end{cases}
\end{equation}
where $\mu_{RR} = 0.91(1 - p)\eta / \sigma_Y$ and $\sigma_{RR} = 0.91(1-p) \tau / \sigma_Y$.
\end{theorem}

\begin{theorem} \label{theorem: evalue2}
Suppose that $\text{Var}(Y) = \sigma_Y$ are known,  that $q = 1 - \mathbb{P}(R = 1)$ follows a normal distribution $\mathrm{N}(\mu_q, \sigma_q^2)$, and that $\delta$ follows a normal distribution $\mathrm{N}(\eta, \tau^2)$. Moreover, let $\rho_1 = \sigma_q / \mu_q, \rho_2 = \tau / \eta$. Assume that $\rho_1$ and $\rho_2$ are arbitrarily small, then the 
density function of $V$ can be approximated by
\begin{equation}
f_V(v) = 
\begin{cases}
\frac{1}{\sqrt{2 \pi} \sigma_{RR}} \exp \left\{ - \frac{ \left( \ln \frac{v^2}{2v-1} - \mu_{RR} \right)^2 } { 2 \sigma_{RR}^2} \right\} \left\{ \frac{1}{v} - \frac{1}{v} \frac{1}{2v-1} \right\} & \text{if $RR > 1$}, \\
\frac{1}{\sqrt{2 \pi} \sigma_{RR}} \exp \left\{ - \frac{ \left( \ln \frac{2v-1}{v^2} - \mu_{RR} \right)^2 } { 2 \sigma_{RR}^2} \right\} \left\{ \frac{1}{v} - \frac{1}{v} \frac{1}{2v-1} \right\} & \text{if $0 < RR < 1$} ,
\end{cases}
\end{equation}
where $\mu_{RR} = 0.91\mu_q \eta / \sigma_Y$ and $\sigma_{RR} = 0.91(\mu_q^2 \tau^2 + \eta^2 \sigma_q^2 + \sigma_q^2 \tau^2)^{1/2} / \sigma_Y$.
\end{theorem}

\begin{theorem} \label{theorem: evalue3}
Suppose that $\sigma_Y$ follows an inverse-gamma distribution $\mathrm{IG}(\alpha, \beta)$, that $q = 1 - \mathbb{P}(R = 1)$ follows a normal distribution $\mathrm{N}(\mu_q, \sigma_q^2)$, and that $\delta$ follows a normal distribution $\mathrm{N}(\eta, \tau^2)$. Let $\rho_1 = \sigma_q / \mu_q, \rho_2 = \tau / \eta$, and $\rho_3 = (\mu_q^2 \tau^2 + \eta^2 \sigma_q^2 + \sigma_q^2 \tau^2)^{1/2} / (\mu_q \eta)$. Assume that $\rho_1, \rho_2$, and $\rho_3$ are arbitrarily small, then the 
density function of $V$ can be approximated by
 
\begin{equation}
f_V(v) = 
\begin{cases}
\frac{\beta_V^{\alpha}}{\Gamma(\alpha)} \exp \left\{ - \beta_V \ln \frac{v^2}{2v-1} \right\} \left( \ln \frac{v^2}{2v-1} \right)^{\alpha-1} \left\{ \frac{1}{v} - \frac{1}{v} \frac{1}{2v-1} \right\} & \text{if $RR > 1$}, \\
\frac{\beta_V^{\alpha}}{\Gamma(\alpha)} \exp \left\{ - \beta_V \ln \frac{2v-1}{v^2} \right\} \left( \ln \frac{2v-1}{v^2} \right)^{\alpha-1} \left\{ \frac{1}{v} - \frac{1}{v} \frac{1}{2v-1} \right\} & \text{if $RR < 1$}, 
\end{cases}
\end{equation}
where $\beta_{V} =  \mu_q \eta / (0.91\beta) $.
\end{theorem}

%\begin{theorem}\label{theorem: evalue4}
%Suppose that the propensity score $p$ follows a uniform distribution $U(0, 1)$, that the variance of conversions $\sigma^2$ follows a inverse gamma distribution $IG(\alpha, \beta)$, and that the sensitivity parameter $\delta$ follows a normal distribution $N(0, \tau^2)$. Then the distribution function of $\mu_{missing}$ is
%\begin{equation}
%f(x) = 
%\frac{\sqrt{2\beta}}{2 \tau} \Gamma \left(0, \frac{\sqrt{2 \beta}|x|}{\tau } \right).
%\end{equation}
%The density function of E-value is then given by
%
%\begin{equation}
%f_V(v) =
%\begin{cases}
%\frac{\sqrt{2 \beta}}{1.82 \tau} \left(\frac{1}{v} - \frac{1}{v}\frac{1}{2v-1} \right) \Gamma \left(0, \frac{ \sqrt{2 \beta} }{ 0.91 \tau \ln \frac{v^2}{2v-1}} \right) & \text{if $RR > 1$} ,\\
%\frac{\sqrt{2 \beta}}{1.82 \tau} \left(\frac{1}{v} - \frac{1}{v}\frac{1}{2v-1} \right) \Gamma \left(0, \frac{ \sqrt{2 \beta} }{ 0.91 \tau \ln \frac{2v-1}{v^2}} \right) & \text{if $RR < 1$},
%\end{cases}
%\end{equation}
%where $\Gamma(0, x) = \int_x^{\infty} t^{-1} e^{-t} ~ dt$ is the incomplete gamma function. 
%
%\end{theorem}

\section{Results on Real and Simulated Data}

To empirically demonstrate the advantages and limitations of techniques presented in section \ref{sec: method}, results on simulated data with empirically grounded properties are presented in section \ref{sec: simu}. We supplement this with a case-study on our motivating application in section \ref{sec: data}. Our objective is to compare the quality of uncertainty quantification and downstream conclusions drawn relative to \textit{asymptotic} estimators of uncertainty that rely on large sample theory. E-value variance is estimated using (1) a Taylor Series estimate of uncertainty %WU, take a look at the chapters on missing data I sent you; there is a reference there for using the delta method for variance approximation, cite that here. % 
(see chapter 5 of \cite{LittDC12})
and (2) Poisson Sampling Theory \cite{Kott06} %WU, add a reference to the paper for Poisson Sampling here%.
. These estimates are utilized in the formulation from \cite{VandD17} to construct asymptotic uncertainty intervals. For our proposed techniques, credible intervals for the E-value are constructed via posterior simulation.

\subsection{Simulation} \label{sec: simu}

%\subsection{Outline}
%
%\begin{itemize}
%
%\item Describe simulation setting (1) there is no misspecification, i.e. $\delta$ is zero plus noise and (2) there is misspecification i.e. $\delta$ is nonzero plus noise; (Any suggestions on the simulation setting?)
%and the purpose of conducting simulation: comparing proposed methods with existing approaches. 
%
%\item Report (1) CI Coverage Rate and (2) False Positive Rate 
%
%% Wu, I think you can explore: (1) CI Coverage and (2) False positive rate those are the two that make the most sense.
%
%\end{itemize}

\begin{figure}[ht]
\centering
\includegraphics[width=0.8\textwidth]{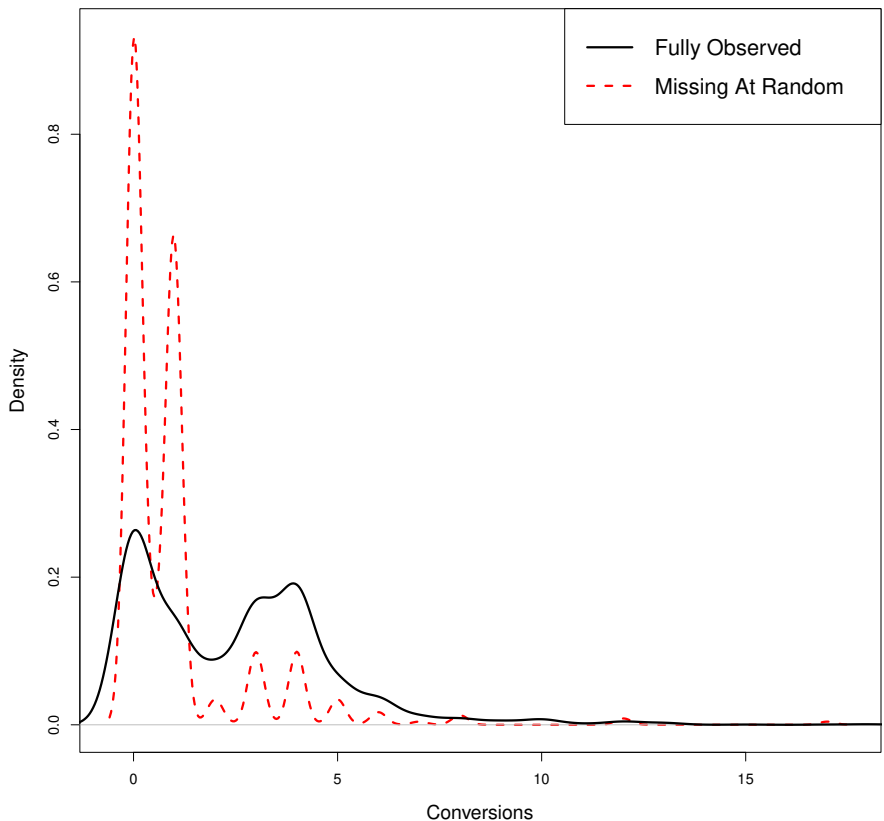}
\caption{Distribution of conversion counts when outcome data is fully observed versus when it is missing at random.}
\label{fig: convs_density}
\end{figure}

In order to evaluate our methodology we utilize simulated data that mimics our motivating application with known parameters. Each simulated data set contains $i = 1,\ldots, 2500$ independent units of observation. For each unit $i$, the outcomes $Y_{i}$, covariates $\bX_{i}$, estimated propensity scores $\hat{\pi}(\bX_{i})$ and related missingness status $R_i \sim \mathrm{Bern} ( \hat{\pi}(\bX_{i}))$ are simulated via sub-sampling from \textit{Facebook} data. The sensitivity parameter estimates $\boldsymbol\delta_j ( j = 1, \ldots , 15) $ are generated from a bivariate normal distribution $
\boldsymbol\delta_j \sim N_2 \left( \begin{pmatrix} 0 \\ 0 \end{pmatrix}, \begin{pmatrix} 0.0025 & 0.0004\\0.0004 & 0.0025 \end{pmatrix} \right).
$ An example of the simulated data for conversions is presented in Figure \ref{fig: convs_density}. 
%with mean $(0, 0)^{\prime}$ and covariance matrix $\begin{bmatrix} 0.0025 & 0.0004\\0.0004 & 0.0025 \end{bmatrix}$. 

Therefore in this simulation study, the \text{true} sensitivity parameter is on average zero but observed with some noise. We simulate $T=10,000$ data sets and assess whether the uncertainty intervals from the four approaches correctly fail to reject the baseline E-value $(1-\alpha)\%$ of the time (i.e., the \textit{coverage} rate of the baseline under the interval) where $\alpha$ is the Type-I error rate. This is supplemented by an analysis of interval widths.

The coverage rates for the intervals are given in Table \ref{table: tb1}.  
As the sample size of observed data increases, the confidence intervals from asymptotic approaches come to attain expected coverage rates. On the other hand, although the coverage rate for the subjective Bayesian approach is marginally lower than the desired level, it is robust against different sample sizes. Therefore, the subjective Bayesian approach is applicable when the observed data size is relatively small, a concern often encountered in our application. The objective Bayesian approach is also robust against different observed data sizes, but it may be too conservative particularly when large samples are available. At small sample sizes it also out-performs the asymptotic approaches and may be a good fallback option if the optimization needed for the subjective Bayesian approach is infeasible.

\begin{table}[H] 
\caption{Coverage rate of $95\%$ uncertainty interval for the E-value with different sample sizes of training data (denoted by $k$).}
\label{table: tb1}
\centering
\begin{tabular}{ccccc} \toprule
    {$k$} & Taylor Series & Poisson Sampling & Subjective Bayesian & Objective Bayesian \\ \midrule
    1{$\times$} & 1.0000 & 1.0000 & 0.9448 & 0.9810  \\
    3{$\times$} & 1.0000 & 1.0000 & 0.9448 & 0.9810  \\
    6{$\times$} & 0.9996 & 0.9999 & 0.9448 & 0.9810  \\
    9{$\times$} & 0.9973 & 0.9988 & 0.9448 & 0.9810  \\
    12{$\times$} & 0.9910 & 0.9943 & 0.9448 & 0.9810  \\
    15{$\times$} & 0.9811 & 0.9865 & 0.9448 & 0.9810  \\
    18{$\times$} & 0.9656 & 0.9742 & 0.9448 & 0.9810  \\ \bottomrule
\end{tabular}
\end{table}

We compare the average width of uncertainty intervals from the four types of methods with respect to different sample sizes of observed data in Table \ref{table: tb2}. %WU Add a table reference%. 
The average width of the uncertainty intervals from the two asymptotic approaches are orders of magnitude larger when the sample size is relatively small (close to the motivating application). The average width decreases rapidly when the sample size of observed data increases. By comparison the average widths of the credible intervals from the two Bayesian approaches are stable across the sample size of the observed data. This makes them more reliable for uncertainty quantification than their asymptotic counterparts particularly for problems where sample sizes are unpredictable (like in the motivating application).

\begin{table} [H] 
\caption{Average width of $95\%$ uncertainty interval for the E-value with different sample sizes of training data (denoted by $k$).}
\label{table: tb2}
\centering
\begin{tabular}{ccccc} \toprule
    {$k$} & Taylor Series & Poisson Sampling & Subjective Bayesian & Objective Bayesian \\ \midrule
    1{$\times$} & $7.8255\times10^5 $ & $2.1685\times10^9$ & 0.1380 & 0.1504  \\
    3{$\times$} & 88.4024 & $1.3056 \times 10^3$ & 0.1469 & 0.1601  \\
    6{$\times$} & 2.0216 & 3.8857 & 0.1513 & 0.1649  \\
    9{$\times$} & 1.0315 & 1.2356 & 0.1531 & 0.1670  \\
    12{$\times$} & 0.7996 & 0.8909 & 0.1542 & 0.1682  \\
    15{$\times$} & 0.6824 & 0.7407 & 0.1548 & 0.1689  \\
    18{$\times$} & 0.6078 & 0.6506 & 0.1553 & 0.1694  \\ \bottomrule
\end{tabular}
\end{table}

\subsection{Motivating Application: \textit{Facebook}} \label{sec: data}
\textit{Facebook} systems often rely on inference from missing outcome data (e.g., whether an item was purchased may not always be observed) in order to deliver an engaging and enjoyable experience on the platform. In these settings, IPW methods may be utilized to ensure that bias from self-selection can be eliminated in the estimation of population averages which play a crucial role in many systems. 

We apply the proposed method of Sensitivity Analysis to study the robustness of these IPW estimates of population averages. There exists a risk of the missingness mechanism being MNAR due to misspecified \textit{weighting} models being used in the construction of estimates. For each observation in the data, we may have the following information:
\begin{itemize}

\item \texttt{conversions}: Number of a certain type of events from a single user (e.g., Purchases).

\item \texttt{event name}: Type of the event, taking 14 levels including \textit{Start Trial, Submit Application, Contact, Add To Cart, Add Payment Information, Search, View Content, Complete Registration, Initiate Checkout, Purchase, Schedule, Subscribe, Lead, Add To Wishlist}.

\item \texttt{propensity scores}: Estimated propensity score of being missing.

\end{itemize}

The outcome of interest here are the conversion events. Respecting user data privacy and compliance with regulatory reform, Facebook utilizes aggregated conversions as the approximate ground truth values for the population mean (via the \textit{Aggregated Events Measurement} or AEM system). In this work, we take the differences between IPW estimated average conversions and approximate ground truth averages as the estimates of sensitivity parameters for various types of events.

\begin{figure}[ht]
\centering
\includegraphics[width=0.8\textwidth]{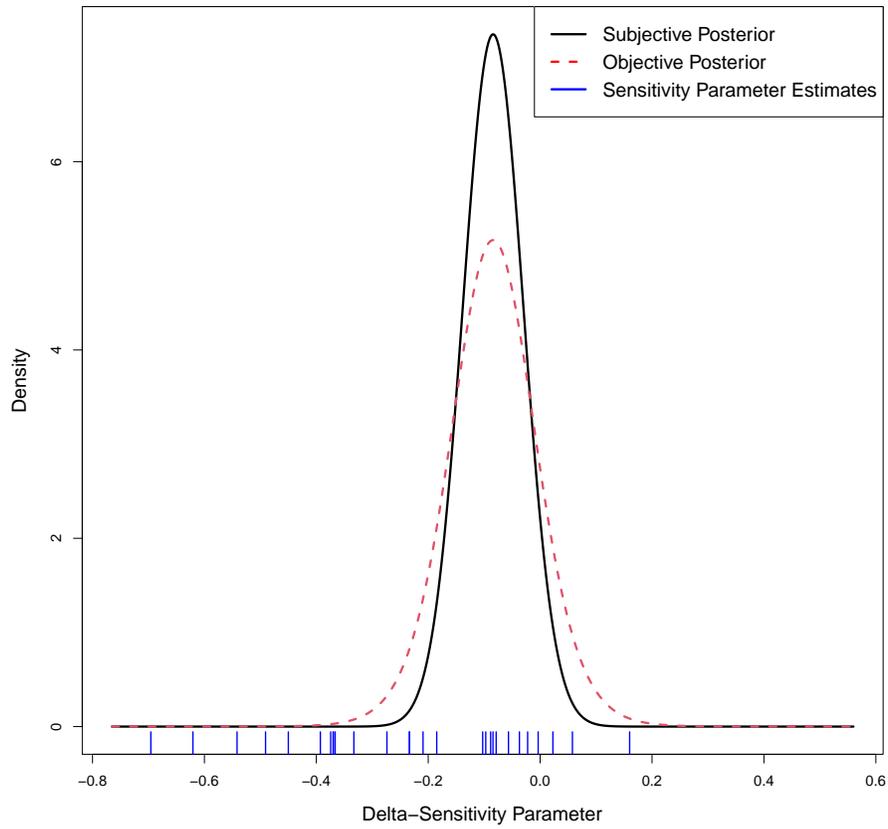}
\caption{Estimated sensitivity parameters and posterior distributions of the sensitivity parameters under both subjective and objective priors}
\label{fig: density2}
\end{figure}

To apply the proposed Bayesian approaches, we obtain the marginal posterior distribution of the sensitivity parameter based on the estimates and our two possible prior specifications. We utilize direct posterior sampling \cite{GelfHRS90} %WU, add a reference to the gelfand paper on gibbs sampling here%
 to generate values of the sensitivity parameter from its implied distribution and calculate the corresponding E-values. The distributions of the sensitivity parameters under both prior choices are visualized in Figure \ref{fig: density2}.

\newpage

The $95\%$ asymptotic confidence intervals and the $95\%$ Bayesian credible intervals are summarized in Table \ref{table: tb3} and \ref{table: tb4}. %WU, use references for these tables as well%.

\begin{table}[H]
\caption{$95\%$ uncertainty interval for E-value for different types of events from in-application data}
\label{table: tb3}
\centering
\resizebox{\textwidth}{!}{
\begin{tabular}{ccccc} \toprule
    Event Name & Taylor Series & Poisson Sampling & Subjective Bayesian & Objective Bayesian \\ \midrule
    Start Trial & (1.9825, 2.8795)              & (1.9816, 2.8616)   & (1, 1.6215) & (1, 1.7806)  \\
    Submit Application	& (1.6015, 2.5573) & (1.5893, 2.5514)    &	(1, 1.6029) &	(1, 1.7557)  \\
    Contact	 & (1.2941, 2.1228)                 & (1.2890, 2.1112)   & (1, 1.5976) &	(1, 1.7487) \\
    Add To Cart &	(1.9844, 2.2117)     &	 (1.9836, 2.2081)   &	(1, 1.6023) &	(1, 1.7549) \\
    Add Payment Info & (2.0237, 3.0502) & (2.0124, 3.0424)   & (1, 1.6076) &	(1, 1.7620) \\
    Search & (2.4783,	2.8783)                  &	(2.4739, 2.8751)   &	(1, 1.5951) &	(1, 1.7453) \\
   View Content	& (2.1910,	2.2559)          & (2.1902, 2.2554)   &	(1, 1.5961) &	(1, 1.7467) \\
   Complete Registration	& (1.5749, 1.8631) & (1.5726, 1.8599) & (1, 1.6059) & (1, 1.7597) \\
   Initiate Checkout &	(1.4163, 1.8987)    &	(1.4135, 1.8921)	& (1, 1.6032) &	(1, 1.7561) \\
   Purchase &	(1.9800, 2.1216)            &	(1.9788, 2.1201)	& (1,	1.6074) &	(1, 1.7618) \\
   Schedule &	(1.5842, 2.5758)            &	(1.5844, 2.5547)	& (1,	1.6083) &	(1, 1.7629) \\
   Subscribe &	(1.3760, 2.6290)            &	(1.3524, 2.6275)	& (1,	1.6128) &	(1, 1.7690) \\
   Lead	&      (1.9731, 2.1028)            &	(1.9722, 2.1012)	& (1,	1.6057) &	(1, 1.7594) \\
   Add To Wishlist &	(1.3687, 2.2694)   &	(1.3522, 2.2675)       & (1, 1.6018) &	(1, 1.7543) \\ \bottomrule
\end{tabular}}
\end{table}

We find that the credible intervals from Bayesian approaches are more stable across different types of events when compared with their counterparts from asymptotic techniques. This is consistent with the simulation study where the properties of Bayesian credible intervals are not affected by the sample size of the observed data. 
%Although all the uncertainty intervals for E-values from IAW include one, there are two inconsistencies from CAPI. The confidence intervals from the Taylor series approximation of \textit{Schedule} and \textit{Add To Wishlist} do not include 1 while the credible intervals from the Bayesian methods do. We prefer the Bayesian approach since they are more stabile across event types and that they have better performance when the sample size of observed data is relatively small (which is the case for \textit{Schedule} and \textit{Add To Wishlist}).
Moreover, there exist notable differences in conclusions between the results from the asymptotic and Bayesian approaches, e.g., for \textit{Search} and  \textit{Purchase} events as in Table \ref{table: tb4}. For \textit{Search}, the confidence intervals from the two asymptotic approaches do not include the null value (E-value $= 1$) while their Bayesian variants result in conservative conclusions. For \textit{Purchase}, the confidence interval from the Taylor series approximation of the uncertainty is solely significant while the others remain conservative. We propose relying on the Bayesian approach here given the sample size of observed data and the findings of the performance from simulation.

\begin{table}[H]
\caption{$95\%$ uncertainty interval for E-value for different types of events from advertiser server data}
\label{table: tb4}
\centering
\resizebox{\textwidth}{!}{
\begin{tabular}{ccccc} \toprule
    Event Name & Taylor Series & Poisson Sampling & Subjective Bayesian & Objective Bayesian \\ \midrule
    Start Trial & (1, 1.5646) & (1, 1.2293) & (1, 1.1064) & (1, 1.1254)  \\
    Submit Application	& (1,	1.7366) & (1, 1.2706) &	(1, 1.0995) &	(1, 1.1171)  \\
%   Contact	 & (1, 1.92691) & (1, 1.91222) & (1, 1.19653) &	(1, 1.23630) \\
    Add To Cart &	(1, 1.2657) &	(1, 1.5144) &	(1, 1.1412) &	(1, 1.1672) \\
    Add Payment Info & (1, 1.7273) & (1, 1.3366) & (1, 1.1425) &	(1, 1.1688) \\
    Search & (1.2284,	1.4240) &	(1.0532, 1.5202) &	(1, 1.1496) &	(1, 1.1773) \\
   View Content	& (1,	1.1658) & (1, 1.2253) &	(1, 1.1420) &	(1, 1.1681) \\
   Complete Registration	& (1, 1.3306) & (1, 1.0898) &	(1, 1.1282) &	(1, 1.1515) \\
   Initiate Checkout &	(1, 1.3566) &	(1, 1.5983)	& (1, 1.1372) &	(1, 1.1624) \\
   Purchase &	(1.0479, 1.1843) &	(1, 1.2697)	& (1,	1.1297) &	(1, 1.1532) \\
%   Schedule &	(1.22321, 1.55610) &	(1, 1.72039)	& (1,	1.18563) &	(1,	1.22285) \\
   Subscribe &	(1, 1.4992) &	(1, 1.1472)	& (1,	1.1156) &	(1, 1.1363) \\
   Lead	&      (1, 1.1922) &	(1, 1.3087)	& (1,	1.1304) &	(1, 1.1541) \\
   Add To Wishlist &	(1, 1.3619) &	(1, 1.4663) &	(1, 1.1423) &	(1, 1.1685) \\ \bottomrule
\end{tabular}}
\end{table}

\section{Discussion} \label{sec: disc}

%\subsection{Outline}
%
%
%\begin{itemize}
%
%\item Review of what we did: Propose two Bayesian Sensitivity Analysis Approaches. Conduct Bayesian Sensitivity Analysis in Signal Loss Problem at scale. 
%
%\item Future work topics: we only considered MAR and non-MAR in this work. What about different MNAR patterns? What can we do?
%
%
%\end{itemize}

To ensure the robustness of inferences from missing data, this paper introduces methods for Sensitivity Analysis by extending the concept of the E-value under the Bayesian paradigm. This conceptualization rests upon sensitivity parameters as differences between noisy benchmarks (e.g., privacy-centric aggregates such as those from Google's \textit{Privacy Sandbox}) and their estimates learned from partially missing unit level outcomes. Treating these differences as data and leveraging priors over the sensitivity parameters, helps to quantify the robustness of inference against MAR violations under the Bayesian framework. We demonstrated performance gains on real and simulated data motivated by applications at Facebook where missing unit-level outcomes are omnipresent.

This paper makes several novel contributions to the field of Sensitivity Analysis for missing data. To the best of our knowledge, we are the first to study the distribution function of the E-value for missing data under a Bayesian framework. We propose two novel Bayesian characterizations to derive the posterior distribution of the sensitivity parameters and consequently the distribution function of the E-value. Our theoretical findings are supplemented by the empirical benefits of this approach. We demonstrate improvements in uncertainty quantification while reducing the reliance on asymptotic guarantees (which may by implausible for the large scale assessment of conclusions from missing data).

The assumptions we make and challenges we encounter in our motivating application lay the foundations for future work. First, our proposed methods rely on improving confidence in conclusions by pooling information. It is natural to borrow strength from sensitivity parameter estimates of outcomes that are similar (as an example closely related conversion events). For more general types of information pooling (e.g., across different categories of events that may not be strongly influenced by each other) added flexibility in assumptions is needed. Second, our current technique is restricted to cross-sectional analyses; extensions to Sensitivity Analysis of longitudinal data with missingness will require understanding how sensitivity parameters can be effectively estimated over time. In this respect, hierarchical priors \cite{Berl96, WiklBC98} may be leveraged when similarity exists both within and across subgroups of sensitivity parameter estimates. These concepts offer promising avenues of future work that we intend to explore.

\section*{Acknowledgements}

The authors are grateful to Richard Mudd, Michael Gill and Qing Feng for their valuable advice and insightful suggestions that are reflected throughout this manuscript. 

\appendix
\section{Appendix}\label{sec:app}

\subsection{Proof of Proposition \ref{proposition: convex}}

%Then the marginal likelihood is the ratio of joint density $f(\boldsymbol\delta_1, \ldots, \boldsymbol\delta_n, \delta, \bSigma)$ to the posterior distribution $\pi(\delta, \bSigma | \boldsymbol\delta_1, \ldots, \boldsymbol\delta_n)$, 
%\begin{align*}
%m(\boldsymbol\delta_1, \ldots, \boldsymbol\delta_n) 
%&=
%    (2\pi)^{-n} \frac{|\bPsi|^{\nu/2} }{2^{\nu} \Gamma_2(\nu/2)}\frac{2^{\tilde{\nu}} \Gamma_2(\tilde{\nu}/2)}{|\tilde{\bPsi}|^{\tilde{\nu}/2} } \\
%    &=
%    \frac{1}{\pi^n} \frac{|\bPsi|^{\nu/2}}{|\tilde{\bPsi}|^{\tilde{\nu}/2}}\frac{ \Gamma_2(\tilde{\nu}/2)}{\Gamma_2(\nu/2)}
%\end{align*}

Our goal is to determine conditions for the convexity of the objective function in order to ensure that its optimization with respect to the parameters $\delta_{0}, \nu$ and $\bPsi$ can be readily handled. From the motivating application $\bPsi$ is assumed to be  $2\times 2$ here. 

Taking the negative logarithm of the marginal likelihood function (\ref{equation: likelihood}) yields the objective function with respect to $\delta_{0}, \nu$ and $\bPsi$,
\begin{align*}
\mathcal{L} (\delta_{0}, \nu,\bPsi) 
&=
m \log(\pi)  - \frac{\nu}{2} \log |\bPsi| + \frac{\tilde{\nu}}{2} \log | \tilde{\bPsi} | + \log \frac{ \Gamma_2(\nu/2)}{\Gamma_2(\tilde{\nu}/2)} 
\end{align*}

Taking the first derivative with respect to $\bPsi$,
\begin{align*}
\frac{\partial \mathcal{L} (\delta_{0}, \nu,\bPsi) }{\partial \bPsi} &= - \frac{\nu}{2}  \bPsi ^{-1} + \frac{\tilde{\nu}}{2} \tilde{\bPsi} ^{-1}\stackrel{\text{set}}{=} 0,
\end{align*}
we get
$$\bPsi ^ {*} = \frac{\nu}{m} \left( \bS + \frac{ m }{ m  + 1}(\bar{\boldsymbol\delta} - \delta_0 \bone)(\bar{\boldsymbol\delta} - \delta_0 \bone)^{\prime} \right).$$

If the objective function is convex with respect to $\bPsi$, then $\bPsi ^ {*}$ will be its global minimizer. 
%and also show that the objective function is convex in $\delta_0$ given certain values of $\nu$. 
We demonstrate that $f(\bPsi) = - \frac{\nu}{2} \log |\bPsi| + \frac{\tilde{\nu}}{2} \log | \tilde{\bPsi} |$ is convex for certain fixed values of $\nu$ by considering an arbitrary line given by $\bPsi + t \bV$, where $\bPsi$ and $ \bV$ are positive definite matrices.

Define $g(t) = - \frac{\nu}{2} \log |\bPsi + t \bV| + \frac{\tilde{\nu}}{2} \log | \tilde{\bPsi} + t \bV |$ such that $\bPsi + t \bV$ and $\tilde{\bPsi} + t \bV$ are positive definite matrices. Since $\bPsi$ and $\tilde{\bPsi}$ are positive definite, there exist $\bPsi^{1/2}$ and $\tilde{\bPsi}^{1/2}$ such that $\bPsi = \bPsi^{1/2}\bPsi^{1/2}$ and $\tilde{\bPsi} = \tilde{\bPsi}^{1/2}\tilde{\bPsi}^{1/2}$. Hence,

\begin{align*}
g(t) &=  - \frac{\nu}{2} \log |\bPsi + t \bV| + \frac{\tilde{\nu}}{2} \log | \tilde{\bPsi} + t \bV | \\
&=
 - \frac{\nu}{2} \log | \bPsi^{1/2}\bPsi^{1/2} + t \bPsi^{1/2}\bPsi^{-1/2} \bV \bPsi^{-1/2}\bPsi^{1/2} | \\
 &+
 \frac{\nu+m}{2} \log |\tilde{\bPsi}^{1/2}\tilde{\bPsi}^{1/2} + t \tilde{\bPsi}^{1/2}\tilde{\bPsi}^{-1/2}\bV\tilde{\bPsi}^{-1/2}\tilde{\bPsi}^{1/2} | \\
 &=
 - \frac{\nu}{2} \log | \bPsi^{1/2} (\bI + t \bPsi^{-1/2} \bV \bPsi^{-1/2} )\bPsi^{1/2} |  \\
 &+ 
  \frac{\nu+m}{2} \log |\tilde{\bPsi}^{1/2} ( \bI + t \tilde{\bPsi}^{-1/2}\bV\tilde{\bPsi}^{-1/2})\tilde{\bPsi}^{1/2} | \\
 &=
 - \frac{\nu}{2} \left( \log | \bPsi | + \log | \bI + t \bPsi^{-1/2} \bV \bPsi^{-1/2} | \right) \\ 
 &+ 
 \frac{\nu+m}{2} \left( \log |\tilde{\bPsi} | + \log |  \bI + t \tilde{\bPsi}^{-1/2}\bV\tilde{\bPsi}^{-1/2} | \right) \\
 &=
  - \frac{\nu}{2} \left\{ \log | \bPsi | + \log ( 1+ t \lambda_1) + \log ( 1+ t \lambda_2)  \right\} \\ 
  &+ 
  \frac{\nu+m}{2} \left\{ \log |\tilde{\bPsi} | + \log (1+ t \eta_1) + \log (1 + t \eta_2) \right\},
\end{align*}
where $\lambda_1, \lambda_2$ are eigenvalues of $ \bI + t \bPsi^{-1/2} \bV \bPsi^{-1/2}$ and $\eta_1, \eta_2$ are eigenvalues of $ \bI + t \tilde{\bPsi}^{-1/2}\bV\tilde{\bPsi}^{-1/2} $. Since $\bI + t \bPsi^{-1/2} \bV \bPsi^{-1/2}$ and $ \bI + t \tilde{\bPsi}^{-1/2}\bV\tilde{\bPsi}^{-1/2} $ are also positive definite matrices,

\begin{align*}
g^{\prime}(t) 
&=
-\frac{\nu}{2} \left( \frac{\lambda_1}{1 + t\lambda_1} + \frac{\lambda_2}{1 + t\lambda_2} \right) + \frac{\nu + m}{2} \left( \frac{\eta_1}{1 + t\eta_1} + \frac{\eta_2}{ 1 + t\eta_2 } \right) \\
&=
- \frac{\nu}{2}  \left\{ \left( \frac{1}{t + \lambda_1^{-1}} + \frac{1}{ t + \lambda_2^{-1}} \right) - \left( \frac{1}{ t + \eta_1^{-1} } + \frac{1}{ t + \eta_2^{-1} } \right)\right\} \\
&+ 
\frac{m}{2} \left( \frac{1}{ t + \eta_1^{-1} } + \frac{1}{ t + \eta_2^{-1} } \right),
\end{align*}
and
\begin{align*}
g^{\prime \prime}(t) &= \frac{\nu}{2} \left\{ \frac{\lambda_1^2}{(1 + t\lambda_1)^2} + \frac{\lambda_2^2}{(1 + t\lambda_2)^2} \right\} -\frac{\nu + m}{2} \left\{ \frac{\eta_1^2}{(1 + t\eta_1)^2} + \frac{\eta_2^2}{(1 + t\eta_2)^2} \right\} \\
&=
\frac{\nu}{2} \left[ \left\{ \frac{1}{(t + \lambda_1^{-1})^2} + \frac{1}{(t + \lambda_2^{-1})^2} \right\} - \left\{ \frac{1}{(t + \eta_1^{-1})^2} + \frac{1}{(t + \eta_2^{-1})^2} \right\} \right] \\
& - \frac{m}{2}\left\{ \frac{1}{(t + \eta_1^{-1})^2} + \frac{1}{(t + \eta_2^{-1})^2} \right\}.
\end{align*}
Since
$\tilde{\bPsi} = \bPsi + \bS + \frac{ m }{m+1}(\bar{\boldsymbol\delta} - \delta_0 \bone)(\bar{\boldsymbol\delta} - \delta_0 \bone)^{\prime}$, it is easy to verify that $\lambda_1 > \eta_1$ and $\lambda_2 > \eta_2$, which gives a range of values of $\nu$ over which the objective function is convex in $\bPsi$ as,
 
\begin{align}
\nu > m \left[ \left\{ \frac{1}{(t + \lambda_1^{-1})^2} + \frac{1}{(t + \lambda_2^{-1})^2} \right\} - \left\{ \frac{1}{(t + \eta_1^{-1})^2} + \frac{1}{(t + \eta_2^{-1})^2} \right\} \right]^{-1} \cdot
\end{align}
$$
\left\{ \frac{1}{(t + \eta_1^{-1})^2} + \frac{1}{(t + \eta_2^{-1})^2} \right\}.
$$

%Moreover, the objective function can be taken as a composite function $\mathcal{L} ( \bPsi (\delta_0) )$ when $\nu$ is fixed. It is easy to check that 
%$
%\frac{ \partial }{\partial \delta_0} \frac{ \partial \bPsi }{\partial \delta_0} 
%=
%\frac{2m}{m+1} \bone \bone^{\prime}
%$
%is a positive definite matrix. Therefore, $\bPsi (\delta_0) $ is convex in $\delta_0$. Based on the composition of convex functions \cite{BoydV04}, the objective function is convex in $\delta$ when it is convex and non-decreasing in $\bPsi$ and $\bPsi$ is convex in $\delta_0$. We already showed that the objective function is convex in $\bPsi$ and $\bPsi$ in convex in $\delta_0$, we additionally need that $g^{\prime}(t) \geq 0$ which leads to 
%\begin{align}
%\nu &\leq 
%m \left\{ \left( \frac{1}{ t + \lambda_1^{-1}} + \frac{1}{ t + \lambda_2^{-1}} \right) - \left( \frac{1}{ t + \eta_1^{-1}} + \frac{1}{ t + \eta_2^{-1} } \right)\right\}^{-1} \left( \frac{1}{ t + \eta_1^{-1} } + \frac{1}{ t + \eta_2^{-1} } \right).
%\end{align}

%To motivate intuition further, the plot of the values of objective function versus $\delta_0$ for certain given values of $\Psi^*$ and $\nu^*$ is presented in Figure \ref{fig: convex}.
%
%\begin{figure}[H]
%\centering
%\includegraphics[width=0.8\textwidth]{convexity_delta.eps}
%\caption{Plot of objective function for certain given values of $\Psi^*$ and $\nu^*$}
%\label{fig: convex}
%\end{figure}

%\textcolor{red}{Wu, this section needs revision, from here}
Substituting $\bPsi^*$ into the objective function and optimizing over $\nu$ leads to,
\begin{align*}
\mathcal{L} (\delta_0, \bPsi^*, \nu) 
&=
\nu \log  \frac{m + \nu}{\nu}  + m \log \frac{m + \nu}{m} +  \log \frac{ \Gamma_2(\nu/2)}{\Gamma_2(\tilde{\nu}/2)} + \text{const}. 
\end{align*}
%and
%\begin{align*}
%\frac{\partial \mathcal{L} (\delta, \bPsi^*, \nu) }{\partial \nu}
%&= 
%\log \frac{m+\nu}{\nu} + \frac{\partial}{\partial \nu} \log \frac{ \Gamma_2(\nu/2)}{\Gamma_2(\tilde{\nu}/2)} \\
%\frac{\partial^2\mathcal{L} (\delta,  \bPsi^*, \nu) }{\partial \nu^2}
%&= 
%-\frac{m}{m\nu +\nu^2} + \frac{\partial^2}{\partial \nu^2} \log \frac{ \Gamma_2(\nu/2)}{\Gamma_2(\tilde{\nu}/2)} > 0 \\
%\end{align*}
%
%Note that there is no closed form solution for $\nu$ but it is easily verified that the objective function is convex since its second order derivative with respect to $\nu$ is positive.
%\textcolor{red}{to here. WU, this is where you can add that this cost function is convex with respect to $\delta$}

It warrants mention that this objective function is not convex with respect to $\nu$; it is a monotone decreasing function. Furthermore, there is no closed form solution for its \textit{optimal} value, $\nu^{*}$. Hence $\nu^{*}$ is set to the inflection point of the objective function along values of $\nu$ which offers notions of optimality as discussed in \cite{HannN14}. 
%\textcolor{red}{Wu add a reference to the optimization paper here}.

Substituting $\bPsi^*$ and $\nu^*$ back into the objective function, and collecting the relevant terms we can ascertain convexity conditions with respect to $\delta_0$,

\begin{align*}
\mathcal{L} (\delta_0, \bPsi^*, \nu^*)
&=
\frac{m}{2} \log \left| \bS + \frac{m}{m+1} (\bar{\boldsymbol\delta} - \delta_0 \bone)(\bar{\boldsymbol\delta} - \delta_0 \bone)^{\prime} \right| + \text{const}.
\end{align*}

First notice that $\bS$ is positive definite, so there exists $\bS^{1/2}$ such that $\bS = \bS^{1/2} \bS^{1/2}$. Then,
\begin{align*}
\mathcal{L} (\delta_0, \bPsi^*, \nu^*)
&=
\frac{m}{2} \log \left| \bS^{1/2} \bS^{1/2} + \frac{m}{m+1} (\bar{\boldsymbol\delta} - \delta_0 \bone)(\bar{\boldsymbol\delta} - \delta_0 \bone)^{\prime} \right| + \text{const} \\
&=
\frac{m}{2} \log \left| \bS^{1/2} \left\{ \mathbf{I} + \frac{m}{m+1} \bS^{-1/2}(\bar{\boldsymbol\delta} - \delta_0 \bone)(\bar{\boldsymbol\delta} - \delta_0 \bone)^{\prime} \bS^{-1/2} \right\} \bS^{1/2} \right| + \text{const} \\
&=
\frac{m}{2} \log \left| \bS \right| + \frac{m}{2} \log \left| \mathbf{I} + \frac{m}{m+1}\bS^{-1/2} (\bar{\boldsymbol\delta} - \delta_0 \bone)(\bar{\boldsymbol\delta} - \delta_0 \bone)^{\prime}\bS^{-1/2} \right|  + \text{const} \\
&=
\frac{m}{2} \log \left| \mathbf{I} + \frac{m}{m+1}\bS^{-1/2} (\bar{\boldsymbol\delta} - \delta_0 \bone)(\bar{\boldsymbol\delta} - \delta_0 \bone)^{\prime}\bS^{-1/2} \right|  + \text{const}. 
\end{align*}

Let $\bu = \left(\frac{m}{m+1} \right)^{1/2} \bS^{-1/2} (\bar{\boldsymbol\delta} - \delta_0 \bone) $, i.e., an \textit{affine} transformation of $\delta_{0}$; consequently, it is sufficient to show that the objective function is convex in $\bu$.

\begin{align*}
\mathcal{L} (\bu, \bPsi^*, \nu^*)
&=
\frac{m}{2} \log \left| \mathbf{I} +  \bu \bu^{\prime} \right|  + \text{const} \\
&=
\frac{m}{2} \log (1 + \bu^{\prime} \bu) + \text{const}. \\
\end{align*}
The first and second derivative of the objective function in $\bu$ are given by
\begin{align}
\frac{ \partial \mathcal{L} (\delta_0, \bPsi^*, \nu^*)}{\partial \bu }
&=
 \frac{m \bu^{\prime}}{1 + \bu^{\prime} \bu},  \nonumber \\
\end{align}
and
\begin{align}
\frac{ \partial^2 \mathcal{L} (\delta_0, \bPsi^*, \nu^*)}{\partial \bu  \partial \bu^{\prime}}
&=
\frac{m( 1 -  \bu^{\prime} \bu)}{(1 + \bu^{\prime} \bu)^2}.
\end{align}
Therefore, the objective function is convex in $\bu$ when $\bu^{\prime} \bu \leq 1$, that is 
\begin{align}
\frac{m}{m+1} \left\{ \bS^{-1/2} (\bar{\boldsymbol\delta} - \delta_0 \bone) \right\}^{\prime} \left\{ \bS^{-1/2} (\bar{\boldsymbol\delta} - \delta_0 \bone) \right\} 
&\leq 1 \nonumber \\
(\bar{\boldsymbol\delta} - \delta_0 \bone)^{\prime} \bS^{-1} (\bar{\boldsymbol\delta} - \delta_0 \bone) 
&\leq \frac{m+1}{m}.
\end{align}

\subsection{Proof of Theorem \ref{theorem: evalue1}}
Let $\mu_{missing} = (1 - p) \delta / \sigma_Y$ where $\delta \sim \mathrm{N}(\eta, \tau^2)$, then $\mu_{missing} \sim \mathrm{N}\left( (1-p) \eta/ \sigma_Y, (1-p)^2 \tau^2 / \sigma_Y^2 \right)$. The risk ratio (RR) satisfies $RR \approx \exp ( 0.91 \times \mu_{missing})$ and $RR \sim \text{Log-normal}( \mu_{RR},  \sigma_{RR}^2)$ where $\mu_{RR} = 0.91(1 - p)\eta / \sigma_Y$ and $\sigma_{RR} = 0.91(1-p) \tau / \sigma_Y$. Further, let $V$ denote the E-value. Then,
\begin{equation}
V = 
\begin{cases}
RR + \sqrt{RR(RR-1)} & \text{if $RR > 1$} ,\\
1 & \text{if $RR =1$}, \\
1/RR + \sqrt{1/RR(1/RR-1)} & \text{if $0 < RR < 1$} .
\end{cases}
\end{equation}
Since $\mathbb{P}(RR = 1) = 0$, we will only consider the cases where $RR > 1$ or $0 < RR < 1$. If $RR > 1$, then $RR = V^2 / (2V-1)$ and the density function of $V$ is given by, 
\begin{align*}
f_V(v) &= f_{RR}(v^2 / (2v - 1)) | RR^{\prime} | \\
&=
\frac{1}{\sqrt{2 \pi} \sigma_{RR}} \exp \left\{ - \frac{ \left( \ln \frac{v^2}{2v-1} - \mu_{RR} \right)^2 } { 2 \sigma_{RR}^2} \right\} \left\{ \frac{1}{v} - \frac{1}{v} \frac{1}{2v-1} \right\}.
\end{align*}
Similarly, when $0 < RR < 1$, $RR = (2V-1) / V^2$ and the corresponding density function of $V$ is,
\begin{align*}
f_V(v) &= f_{RR}( (2v - 1) / v^2) | RR^{\prime} | \\
&=
\frac{1}{\sqrt{2 \pi} \sigma_{RR}} \exp \left\{ - \frac{ \left( \ln \frac{2v-1}{v^2} - \mu_{RR} \right)^2 } { 2 \sigma_{RR}^2} \right\} \left\{ \frac{1}{v} - \frac{1}{v} \frac{1}{2v-1} \right\} 
\end{align*}

\subsection{Proof of Theorem \ref{theorem: evalue2}}
Let $\mu_{missing} = q \delta / \sigma_Y$ where $q \sim \mathrm{N}(\mu_q, \sigma_q^2)$ and $\delta \sim \mathrm{N}(\eta, \tau^2)$. By Theorem 2.5, 2.6 and  2.7 in \cite{Aroian47}, under the assumption that $\rho_1 = \sigma_q / \mu_q$ and $\rho_2 = \tau / \eta$ are arbitrarily small, we can \textit{approximate} the distribution of $q\delta$ with a normal distribution with mean $\mu_q \eta$ and variance $\mu_q^2 \tau^2 + \eta^2 \sigma_q^2 + \sigma_q^2 \tau^2$. Under this approximation, we can derive the distribution of the E-value via a similar approach as in proof of Theorem \ref{theorem: evalue1}.

%\cite{WareL03} proposed to approximate the distribution of the product of two normal random variables by a normal distribution. In light of this proposition, we can approximate the distribution of $\mu_{missing}$ by a normal distribution with mean $\mu_q \eta$ and variance $\mu_q^2 \tau^2 + \eta^2 \sigma_q^2 + \sigma_q^2 \tau^2$. Then we are able to derive the distribution of E-value via a similar approach as in proof of Theorem 3.2.

\subsection{Proof of Theorem \ref{theorem: evalue3}}
To prove Theorem \ref{theorem: evalue3}, we apply the normal approximation for the product distribution of $q = 1 - p $ and $\delta$ as in the proof of Theorem \ref{theorem: evalue2} under the assumption that $\rho_1 = \sigma_q / \mu_q$ and $\rho_2 = \tau / \eta$ are arbitrarily small. 
%Moreover, if we assume that $\rho_3 = (\mu_q^2 \tau^2 + \eta^2 \sigma_q^2 + \sigma_q^2 \tau^2)^{-1/2} / (\mu_q \eta)$ is arbitrarily small, 
Thus the characteristic function of $q \delta$ under the normal approximation of the product is $\exp( i \mu_q \eta t  + (\mu_q^2 \tau^2 + \eta^2 \sigma_q^2 + \sigma_q^2 \tau^2) t^2 / 2) $. The characteristic function of $\mu_{missing}$ is therefore, 

\begin{align}
\phi_{\mu_{missing}}(t) &=
\mathbb{E}_{\sigma_Y} \left[ \mathbb{E} \left\{ e^{i t/ {\sigma_Y} q\delta} \bigg | \sigma_Y \right\} \right] \nonumber \\
&=
\int_0^{\infty} \frac{\beta^{\alpha}}{\Gamma(\alpha)} \exp( -\beta y + i \mu_q \eta ty + (\mu_q^2 \tau^2 + \eta^2 \sigma_q^2 + \sigma_q^2 \tau^2) t^2 y^2 / 2) y^{\alpha-1} ~ \text{d}y \nonumber \\
%&=
%\frac{1}{(1 - i \mu_q \eta t / \beta)^{\alpha}}.
\end{align} 
We additionally assume that $\rho_3 = (\mu_q^2 \tau^2 + \eta^2 \sigma_q^2 + \sigma_q^2 \tau^2)^{1/2} / (\mu_q \eta)$ is arbitrarily small, then 
\begin{align}
\phi_{\mu_{missing}}(t) &=
\int_0^{\infty} \frac{\beta^{\alpha}}{\Gamma(\alpha)} \exp( -\beta y + i \mu_q \eta t y) y^{\alpha-1} ~ \text{d}y \nonumber \\
&=
\frac{1}{(1 - i \mu_q \eta t / \beta)^{\alpha}}.
\end{align} 
The distribution of $\mu_{missing}$ is hence approximated by a gamma distribution $\mathrm{G}(\alpha, \mu_q \eta / \beta)$. Let $\beta_{V} =  \mu_q \eta / (0.91\beta) $. Following the concept in the proof of Theorem \ref{theorem: evalue1}, we can derive the approximation of the distribution function of E-value by applying a change of variables twice. Specifically, when $RR> 1$, we have $RR = V^2 / (2V - 1)$ which gives the density function of $V$, 
\begin{align}
f_V(v) &=
f_{RR} \left( \frac{v^2}{2v-1} \right) |RR^{\prime} | \nonumber \\
&=
\frac{\beta_V^{\alpha}}{\Gamma(\alpha)} \exp \left\{ - \beta_V \ln \frac{v^2}{2v-1} \right\} \left( \ln \frac{v^2}{2v-1} \right)^{\alpha-1} \left\{ \frac{1}{v} - \frac{1}{v} \frac{1}{2v-1} \right\}.
\end{align}
Analogously, when $0 < RR < 1$ we have $RR = (2V - 1) / V^2 $ and the density function of $V$ is given by,
\begin{align}
f_V(v) &=
f_{RR} \left( \frac{2v-1}{v^2} \right) |RR^{\prime} | \nonumber \\
&=
\frac{\beta_V^{\alpha}}{\Gamma(\alpha)} \exp \left\{ - \beta_V \ln \frac{2v - 1}{v^2} \right\} \left( \ln \frac{2v-1}{v^2} \right)^{\alpha-1} \left\{ \frac{1}{v} - \frac{1}{v} \frac{1}{2v-1} \right\}.
\end{align}

\bibliographystyle{abbrv}
\bibliography{bsa}

\end{document}